\documentclass[structabstract]{aa}
\usepackage{graphicx}
\usepackage{txfonts}
\usepackage{color}

\begin{document}

\title{Cosmic-ray driven dynamo\\ in the interstellar medium of irregular galaxies}

\titlerunning{CR driven dynamo in the ISM of irregular galaxies}
\authorrunning{H. Siejkowski et al.}

\author{
H. Siejkowski\inst{1}
\and M. Soida\inst{1}
\and K. Otmianowska-Mazur\inst{1}
\and M. Hanasz\inst{2}
\and D.J. Bomans\inst{3}
}

\institute{Astronomical Observatory, Jagiellonian University, ul. Orla 171, 30-244 Krak\'ow, Poland
\and
Toru\'n Centre for Astronomy, Nicolaus Copernicus University, 87-148 Toru\'n/Piwnice, Poland
\and
Astronomical Institute of Ruhr-University Bochum, Univerist\"{a}tsstr. 150/NA7, D-44780 Bochum, Germany
}

\abstract
{Irregular galaxies are usually smaller and less massive than their spiral, S0, and elliptical
counterparts. Radio observations indicate that a magnetic field is present in irregular galaxies
whose value is similar to that in spiral galaxies. However, the conditions in the interstellar
medium of an irregular galaxy are unfavorable for amplification of the magnetic field because of the
slow rotation and low shearing rate.}
{We investigate the cosmic-ray driven dynamo in the interstellar medium of an irregular galaxy. We
study its efficiency under the conditions of slow rotation and weak shear. The star formation is
also taken into account in our model and is parametrized by the frequency of explosions and
modulations of activity.}
{The numerical model includes a magnetohydrodynamical dynamo driven by cosmic rays that is injected
into the interstellar medium by randomly exploding supernovae. In the model, we also include
essential elements such as vertical gravity of the disk, differential rotation approximated by the
shearing box, and resistivity leading to magnetic reconnection.}
{We find that even slow galactic rotation with a low shearing rate amplifies the magnetic field, and
that rapid rotation with a low value of the shear enhances the efficiency of the dynamo.  Our
simulations have shown that a high amount of magnetic energy leaves the simulation box becoming an
efficient source of intergalactic magnetic fields.}
{}


\maketitle
\section{Introduction}

Irregular galaxies have lower masses than typical spirals and ellipticals.  In addition, they have
irregular distributions of the star-forming regions, and rotations that are slower than spiral
galaxies by half an order of magnitude (Gallagher \& Hunter 1984). The rotation curves of irregular
galaxies are non-uniform and have a weak shear.  

Radio observations of magnetic fields in spiral galaxies indicate that their magnetic fields have
strong ordered (1--5~$\mu$G) and random (9--15~$\mu$G) components (Beck 2005). A plausible process
fueling the growth in the magnetic energy and flux of these galaxies is magnetohydrodynamical dynamo
(Widrow 2002; Gressel et al. 2008).  The vital conditions required for the dynamo to effectively
amplify the magnetic field are rapid rotation and shear. In irregular galaxies, both quantities seem
to be too low to initiate efficient dynamo action. In contrast, the observations of magnetic field
in irregular galaxies indicate that these galaxies could have strong and ordered magnetic fields
(e.g., Chy\.zy et al. 2000, 2003; Kepley et al. 2007; Lisenfeld et al.  2004).

The most spectacular radio observations to date of irregulars were those performed for the galaxy
NGC~4449 (Chy\.zy et al. 2000). The total strength of its magnetic field is about $14~\mu$G with a
ordered component reaching locally values of $8~\mu$G. These are similar to the intensities observed
for large spirals. A high number of \ion{H}{ii} regions and slow rotation is also observed with
quite large velocity shear (Valdez-Guti\'errez et al. 2002). The radio observations of \ion{H}{i}
around the galaxy indicate that this object is embedded in two large \ion{H}{i} systems that
counter-rotate with respect to the optical part of this galaxy (Bajaja et al. 1994; Hunter et al.
1998, 1999). In addition to these \ion{H}{i} clouds, NGC~4449 contains an unusual ring of \ion{H}{i}
in the outer part of the optical disk (Hunter et al. 1999). This complicated topology of the
\ion{H}{i} velocity field could help in achieving efficient magnetic field amplification (see
Otmianowska-Mazur et al.  2000).

Chy\.zy et al. (2003) found that two other irregular galaxies, NGC~6822 and IC~10 are also
magnetized. The former has a~very low total magnetic field weaker than 5~$\mu$G, a small number of
\ion{H}{ii} regions, and almost rigid rotation (see Sect.~\ref{sec:observations}). These properties
are directly related to the efficiency of the dynamo process in galaxies (see Otmianowska-Mazur et
al. 2000; Hanasz et al. (2006, 2009) that weakly amplifies the magnetic field in this galaxy. The
irregular IC~10 has a total magnetic field strength that varies between 5 and 15~$\mu$G with no
ordered component. Observations performed by Chy\.zy et al.  (2003) indicate that the total magnetic
field is correlated with the number of \ion{H}{ii} regions. The number of the regions is higher than
in NGC~6822, and the rotation of IC~10 has a partly differential character (see
Sect.~\ref{sec:observations}).  Both conditions lead to more rapid magnetic field amplification than
for NGC~6822. As for NGC~4449, IC~10 is embedded in a~large cloud of \ion{H}{i}, which
counter-rotates with respect to the inner disk (Wilcots \& Miller 1998).

Klein et al. (1993) inferred that the Large Magellanic Cloud (LMC) has a large-scale magnetic field
that has the shape of a~trailing spiral structure, similar to normal spiral galaxies.  It is
possible that the amplification of the magnetic field is connected to the differential rotation of
this galaxy present beyond a certain radius (Klein et al. 1993; Luks \& Rohlfs 1992; Gaensler et al.
2005).

We note that polarized radio emission is detected in the irregular galaxy NGC 1569. This galaxy has
a very high star formation rate (Martin 1998) and exhibits bursts of activity in its past (Vallenari
\& Bomans 1996). The radio observation of this galaxy by Lisenfeld et al. (2004) found that the
galaxy has large-scale magnetic fields in the disk and halo. Furthermore, they found that their data
agree that a convective wind could allow for escape of cosmic-ray electrons in to the halo. These
observations are the main reason for undertaking our CR-driven dynamo calculations in irregular
galaxies.  In addition to Lisenfeld et al. (2004), the radio observations of Kepley et al. (2007)
showed that the large-scale magnetic arms visible in NGC~1569 are aligned perpendicularly to the
disk and that the northern part of the disk of the galaxy is inclined at a different angle.
 
Kronberg et al. (1999) realized that dwarf galaxies (apart from their low masses) could serve as an
efficient source of gas and magnetic fields in the intergalactic medium (IGM) during their initial
bursts of star formation in the early Universe. In the case of star-forming dwarf galaxies, we
expect that the dominant driver of a galactic wind are cosmic rays, in contrast to large spirals,
such as the Milky Way, where the thermal driving is most significant (Everett et al. 2008).
Therefore, we applied the model of the CR-driven dynamo to the interstellar medium and conditions of
an irregular galaxy and try to find how much of the magnetic energy can be expelled from the dwarf
galaxies to the IGM.

Many questions about magnetic field amplification in irregular galaxies remain unresolved.  The
physical explanation of this process is difficult to establish because these galaxies rotate slowly,
almost like a solid body.  In this paper, we check how our model of cosmic-ray driven dynamo, which
effectively describes spiral galaxies (Hanasz et al. 2004, 2006, 2009), can be applied to irregular
galaxies. In the present  numerical experiment we attempt to answer how the model input parameters
observed in irregulars (small gravitational potential, gas density, low rotation, and small shear)
influence the magnetic fields within them. We have not taken into account the magnetic field
possibly injected by stars. We plan to study this in the future. We found that in certain conditions
achievable for irregulars it is possible to have efficient magnetic field amplification.

\section{Observations of irregular galaxies}\label{sec:observations}
\begin{table}[t]
\centering
\caption{Main properties of NGC~4449, NGC~6822, and IC~10}
\begin{tabular}{r|ccc}
\hline \hline
 & NGC~4449 & NGC~6822 & IC~10 \\ \hline
 Type & IBm & IB(s)m & dIrr \\
Distance [Mpc] & 4.21 & 0.50 & 0.66 \\
Diameter [kpc] & 5.73 & 1.71 & 1.27 \\
SFR & high & low & high \\
$<\! B_\textrm{tot}\!>$ [$\mu$G] & 5--15 & $\leq 5$& 5--15 \\
$<\! B_\textrm{reg}\!>$ [$\mu$G] & $\simeq 8$ & $\leq 3$ & $\leq 3$ \\ \hline
\end{tabular}
\begin{flushleft}
In the following rows we present: the morphological type of the object (LEDA), the distance
(Karachentsev et al. 2004), the size calculated from $D_{25}$ (LEDA) and distance, the star
formation rate (SFR), and the results of the radio analysis of Chy\.zy et al. (2000, 2003).
\end{flushleft}
\label{tab:props}
\end{table}

To study properties of the irregulars and determine the input parameters for our simulations, we use
observations of NGC~4449, NGC~6822, and IC~10 acquired by Chy\.zy~et~al. (2000, 2003). The main
properties of these objects are presented in Table~\ref{tab:props}.

\begin{figure}[b]
\centering
\includegraphics[width=\columnwidth]{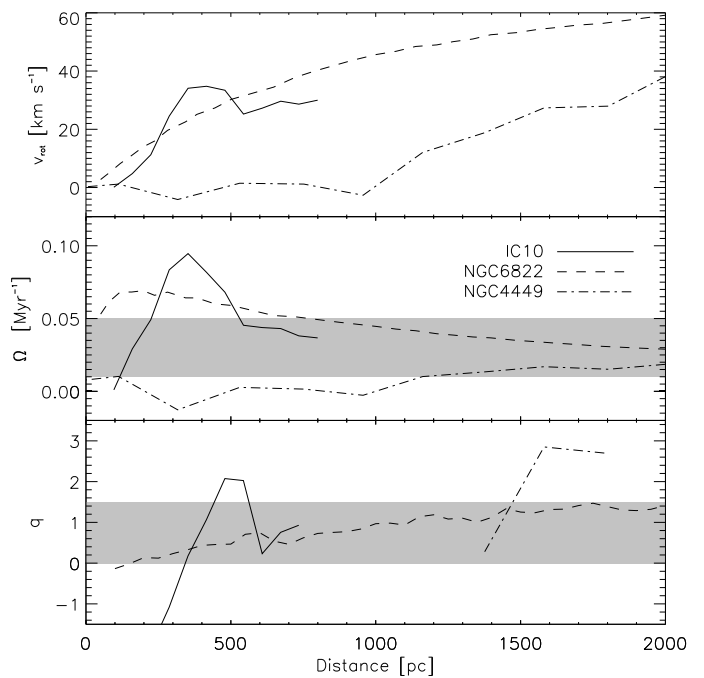}
\caption{Observational rotation characteristics of IC~10, NGC~6822, and NGC~4449. We present, from
top to bottom: the  rotation curves (references in Sect. \ref{sec:observations}), the calculated
angular velocity, and the computed shear parameter $q$ (for details see Sect.~\ref{sec:shearing})
respectively.  The shaded region marks the range of parameters presented in this paper.
\label{fig:shear}}
\end{figure}

From the rotation curves of IC~10 (Fig.~\ref{fig:shear}, top panel, solid line) obtained by Wilcots
\& Miller 1998, NGC~6822 (Fig.~\ref{fig:shear}, top panel, dashed line) obtained by Weldrake et al.
2003, and NGC~4449 (Fig.~\ref{fig:shear}, top panel, dot-dashed line) by Valdez-Guti\'errez et al.
2002, we computed the angular velocity and shearing rate of each galaxy (see Sect.
\ref{sec:shearing}). For the two first galaxies, we use \ion{H}{i} data. In the case of NGC~4449, we
restricted our analysis to the internal region and used H$\alpha$ data, because of its very complex
velocity pattern.

In the velocity pattern of the IC10, we can see a central part with a solid-body rotation, which
flattens to a constant value $v_\mathrm{rot} \simeq 30$~km~s$^{-1}$ at $r=352$~pc.  The NGC~6822
rotation curve is a monotonically increasing function with a square root slope and the highest value
$v_\mathrm{rot} \simeq 60$~km~s$^{-1}$ at $r=5.7$~kpc.  The rotation curve of NGC~4449 is highly
disturbed and reaches a maximum value of $v_\mathrm{rot} \simeq 40$~km~s$^{-1}$ at radius of 2~kpc.

\section{Description of the model} \label{sec:description}
The CR-driven dynamo model consists of the following elements (based on Hanasz et al. 2004, 2006):
\begin{itemize}
\item[(1)] 
The cosmic ray component is a relativistic gas described by a~diffusion-advection transport equation.
Typical values of the diffusion coefficient are
$(3 \div 5) \times 10^{28}~\textrm{cm}^2~\textrm{s}^{-1}$ (see Strong et al. 2007)
at energies of around $1~\textrm{GeV}$, although in our simulations we use reduced values (see Sect.
\ref{sec:model_setup}).
\item[(2)] 
Anisotropic diffusion of CR. Following Giacalone \& Jokipii (1999) and Jokipii (1999), we assume
that the CR gas diffuses anisotropically along magnetic field lines. The ratio of the perpendicular
to parallel CR diffusion coefficients suggested by the authors is $5\%$.
\item[(3)] 
Localized sources of CR. In the model, we apply the random explosions of supernovae in the disk
volume. Each explosion is a localized source of cosmic rays.  The cosmic ray input of
individual SN remnant is $10\%$ of the canonical kinetic energy output ($10^{51}~\textrm{erg}$) and
distributed over several subsequent time steps.
\item[(4)] 
Resistivity of the ISM to enable the dissipation of the small-scale magnetic fields (see Hanasz et
al. 2002 and Hanasz \& Lesch 2003). In the model, we apply the uniform resistivity and neglect the
Ohmic heating of gas by the resistive dissipation of magnetic fields.
\item[(5)] 
Shearing boundary conditions and tidal forces following the prescription by Hawley, Gammie \&
Balbus (1995), are implemented to reproduce the differentially rotating disk in the local
approximation.
\item[(6)] 
Realistic vertical disk gravity following the model by Ferri\`{e}re (1998) modified by reducing the
contribution of disk and halo masses by one order of the magnitude, to adjust the irregular
galaxy environment.
\end{itemize}

We apply the following set of resistive MHD equations:

\begin{eqnarray}
\frac{\partial \rho}{\partial t} &+& \mathbf{\nabla} \cdot (\rho \mathbf{V}) = 0,\\
\frac{\partial e}{\partial t} &+& \mathbf{\nabla} \cdot (e \mathbf{V}) = - p(\mathbf{\nabla} \cdot
\mathbf{V}),\\
\frac{\partial \mathbf{V}}{\partial t} &+& (\mathbf{V} \cdot \mathbf{\nabla}) \mathbf{V} =
- \frac{1}{\rho} \mathbf{\nabla} \left( p + p_{\rm cr} + \frac{B^2}{8 \pi}\right) \nonumber \\
&+& \frac{\mathbf{B} \cdot \mathbf{\nabla B}}{4 \pi \rho} - 2 \mathbf{\Omega} \times \mathbf{V}
+ 2q\Omega^2x \hat{\vec e}_x,\\
\frac{\partial \vec B}{\partial t} &=& \mathbf{\nabla} \times (\vec V \times \vec B) + \eta
\mathbf{\triangle} \vec B, \\
p &=& (\gamma -1)e, \, \; \gamma = 5/3,
\end{eqnarray}
\noindent
where $q= -d \ln\Omega/d \ln R$ is the shearing rate, $R$ is a galactocentric radius, $\eta$
represents the ISM resistivity, $\gamma$ is the adiabatic index of thermal gas, $p_{\rm cr}$ is the
cosmic-ray pressure, and the other symbols have their usual meaning. In the equation of motion, the
term $\nabla p_{\rm cr}$ is included (see Berezinskii et al. 1990). The thermal gas is approximated
by an adiabatic medium.

The cosmic ray component is an additional fluid described by the diffusion-advection equation (see
e.g., Schlickeiser \& Lerche 1985)
\begin{equation}
\frac{\partial e_{\rm cr}}{\partial t} + \mathbf{\nabla} (e_{\rm cr} \vec V) = \mathbf{\nabla}
(\hat{K} \mathbf{\nabla} e_{\rm cr}) - p_{\rm cr}(\mathbf{\nabla} \cdot \vec V) + Q_{\rm SN},
\end{equation}
where $Q_{\rm SN}$ is the source term of the cosmic-ray energy density injected locally from randomly
exploding SN remnants. The cosmic-ray fluid is described by an adiabatic equation of state with
adiabatic index $\gamma_{\rm cr}$:
\begin{equation}
p_{\rm cr} = (\gamma_{\rm cr}-1)e_{\rm cr}, \, \; \gamma_{\rm cr} = 14/9.
\end{equation}
The $\hat{K}$ is an diffusion tensor described by the formula:
\begin{equation}
K_{ij} = K_{\perp} \delta_{ij} + (K_{\parallel} - K_{\perp})n_i n_j, \, \; n_i =  B_i/B,
\end{equation}
adopted following the argumentation of Ryu et al. (2003).

The vertical gravitational acceleration is taken from Ferri\`{e}re (1998). We reduced both
contributions of disk and halo by a factor of~10, the scale length of the exponential disk to $L_D =
2$~kpc, and scale length of halo to $L_H = 1$~kpc. In our computations, we incorporated the formula :
\begin{eqnarray}
\lefteqn{-g_z(R,Z) =
	(1.7 \cdot 10^{-10}\;\textrm{cm s}^{-2}) \frac{R_*^2 + L_H^2}{R + L_H^2}
		\left(\frac{Z}{1\;\textrm{kpc}} \right)} \nonumber \\
&+& (4.4 \cdot 10^{-10}\;\textrm{cm s}^{-2}) \exp\left( -
		\frac{R-R_*}{L_D}\right) \frac{Z}{\sqrt{Z^2 + (0.2\;\textrm{kpc})^2}},
\end{eqnarray}
where $R_*$ is the distance of the origin of the simulation box from the galactic center and $Z$ is
the height above the galactic mid plane.

\section{Model setup and parameters}
\subsection{Model setup} \label{sec:model_setup}

\begin{figure}[t]
\centering
\includegraphics[width=0.9\columnwidth,keepaspectratio=true]{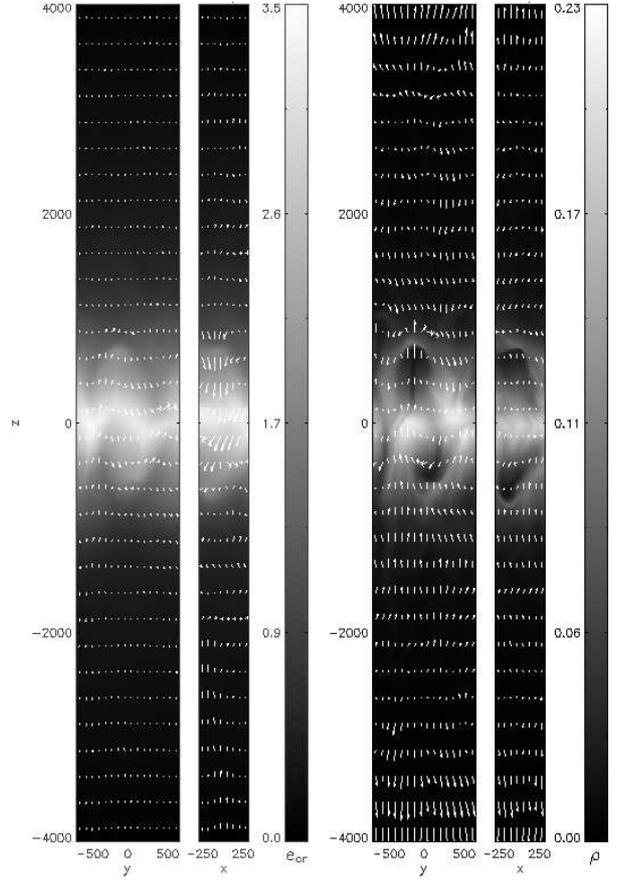}
\caption{Example slices of a domain taken from simulation R.01Q1 at $t=660~\textrm{Myr}$. On the
slices, the Parker loop is produced by cosmic rays from supernovae explosions.
\label{fig:slices}}
\end{figure}

The 3D cartesian domain size is $0.5~\textrm{kpc}\times1~\textrm{kpc}\times8~\textrm{kpc}$ in
$x,y,z$ coordinates corresponding to the radial, azimuthal, and vertical directions, respectively.
The grid size is $20~\textrm{pc}$ in each direction. The boundary conditions are sheared-periodic in
$x$, periodic in $y$, and an outflow in $z$ direction. The domain is placed at the galactocentric
radius $R_* = 2$~kpc. In Fig.~\ref{fig:slices}, we present example slices through the simulation
domain. The left panel shows the CR energy density with the magnetic field vectors and the right
panel shows the gas density with velocity vectors.

The positions of SNe are chosen randomly with a uniform distribution in the $xy$ plane and
a~Gaussian distribution in the vertical direction. The scaleheight of SN explosions in the vertical
direction is $100~\textrm{pc}$, and the CR energy that originates in an explosion is injected
instantaneously into the ISM with a Gaussian radial profile ($r_{\it SN}=50~\textrm{pc}$). In
addition, the SNe activity is modulated during the simulation time by a period $T_p$ and an activity
time $T_a$.

The applied value of the perpendicular CR diffusion coefficient is $K_\perp =
10^3~\textrm{pc}^2~\textrm{Myr}^{-1} = 3 \times 10^{26}~\textrm{cm}^2~\textrm{s}^{-1}$ and the
parallel one is $K_\parallel = 10^{4}~\textrm{pc}^2~\textrm{Myr}^{-1} = 3 \times
10^{27}~\textrm{cm}^2~\textrm{s}^{-1}$. The diffusion coefficients are $10\%$ of realistic values
because of the simulation timestep, which becomes prohibitively short when the diffusion is too
high.

The initial state of the system represents the magnetohydrostatic equilibrium with the horizontal,
purely azimuthal magnetic field with $p_{\it mag}/p_{\it gas} = 10^{-4}$, which corresponds to the
mean value of magnetic field in the simulation box of 5~nG. Magnetic diffusivity $\eta$ is set to be
100~pc$^2$\,Myr$^{-1}$, which corresponds to $3\times10^{25}$~cm$^2$\,s$^{-1}$ in cgs units (Lesch
1993).  The column density of gas is $\varrho_{\it gas} = 6\times 10^{20}~\textrm{cm}^{-2}$ (taken
from observations, see Gallagher \& Hunter 1984) and the initial value of the isothermal sound speed
is set to be $c_{\it iso}=7~\textrm{km}\,\textrm{s}^{-1}$.

\subsection{Model parameters} \label{sec:parameters}

\begin{table}[t]
\centering
\caption{List of models}
\begin{tabular}{l|ccccc}
\hline \hline
Model			& $\Omega$ & $q$ &	$f$ & $T_p$ &	$T_a$ \\
            & [Myr$^{-1}$]
				           & 
							        &  [kpc$^{-2}$~Myr$^{-1}$]
									         & [Myr]
												       & [Myr] \\
\hline
R.01Q1$^a$		  &		0.01 &	1 &	10 &	200 &		20 \\
R.02Q1			    &		0.02 &	1 &	10 &	200 &		20 \\
R.03Q1$^b$		  &		0.03 &	1 &	10 &	200 &		20 \\
R.04Q1			    &		0.04 &	1 &	10 &	200 &		20 \\
R.05Q1$^c$		  &		0.05 &	1 &	10 &	200 &		20 \\
\hline
R.01Q0			    &		0.01 &   0 &	10 &	200 &		20 \\
R.01Q.5			    &		0.01 & 0.5 &	10 &	200 &		20 \\
R.01Q1$^a$		  &		0.01 &   1 &	10 &	200 &		20 \\
R.01R1.5			  &		0.01 & 1.5 &	10 &	200 &		20 \\
R.05Q0			    &		0.05 &   0 &	10 &	200 &		20 \\
R.05Q.5			    &		0.05 & 0.5 &	10 &	200 &		20 \\
R.05Q1$^c$		  &		0.05 &   1 &	10 &	200 &		20 \\
R.05R1.5			  &		0.05 & 1.5 &	10 &	200 &		20 \\
\hline
SF3R.03Q.5		  &		0.03 & 0.5 &	 3 &	200 &		20 \\
SF3R.03Q1		    &		0.03 &   1 &	 3 &	200 &		20 \\
SF10R.03Q.5$^d$ &		0.03 & 0.5 &	10 &	200 &		20 \\
SF10R.03Q1$^b$	&		0.03 &   1 &	10 &	200 &		20 \\
SF30R.03Q.5		  &		0.03 & 0.5 &	30 &	200 &		20 \\
SF30R.03Q1		  &		0.03 &   1 &	30 &	200 &		20 \\
\hline
M10/100			    &		0.03 & 0.5 &	10 &  100 &		10 \\
M20/200$^d$		  &		0.03 & 0.5 &	10 &  200 &		20 \\
M50/100			    &		0.03 & 0.5 &	10 &  100 &		50 \\
M100/200			  &		0.03 & 0.5 &	10 &  100 &	  200 \\
M100/100			  &		0.03 & 0.5 &	10 &  100 &	  100 \\
FIRST				    &		0.03 & 0.5 &  2.5 & 2000 &   50 \\
\hline
\end{tabular}
\begin{flushleft}
Subsequent columns show: the model name, the angular velocity
$\Omega$, the shearing parameter $q$, the frequency of SN explosions $f$, the period of SNe modulation $T_p$,
and the duration of SNe activity in one the period $T_a$. See Sect. \ref{sec:parameters} for details. The
horizontal lines distinguish between different simulation series. Models with the same superscript ($^a$,
$^b$, $^c$, and $^d$) point to the same experiments, but are written for clarity.
\end{flushleft}
\label{tab:models}

\end{table}

We present the results of four simulation series corresponding to different sets of the CR-dynamo
parameters. Details of all computed models are shown in Table~\ref{tab:models}.  The model name
consists of a combination of four letters: R, Q, SF and M followed by a~number.  The letter R means
the angular velocity (rotation), Q~is the shearing rate, SF is the supernova explosion frequency and
M~represents for its modulation during the simulation time, and the numbers determine the value of
the corresponding quantity.  Only the modulation symbol is followed by two numbers, the first
corresponding to the time of the SNe activity and the second to a period of modulation. Values of
the parameters are given in the following units: angular velocity in Myr$^{-1}$, supernova explosion
frequency in kpc$^{-2}$\,Myr$^{-1}$, and the modulation times in Myr. For example, a model named
"R.01Q1" denotes a simulation where $\Omega = 0.01$~Myr$^{-1}$ and $q = 1$, and the name "M50/100"
denotes an experiment with $T_a = 50$~Myr and $T_p = 100$~Myr. The last model in
Table~\ref{tab:models}, named FIRST, points to an experiment, in which only during the first 50~Myr
supernovae are active and after that time CR injection stops.

\section{Results}

\subsection{Shear parameter q obtained from observations} \label{sec:shearing}

The shearing rate parameter $q$ (defined in Sect. \ref{sec:description}) is calculated
numerically from the observational rotation curves using a~second order method
\begin{equation}
q_i = 1 - \frac{R_i}{v_i}\left( \frac{v_{i+1}-v_{i-1}}{R_{i+1}-R_{i-1}} 
+ \frac{1}{2}\frac{v_{i+1}-2v_i+v_{i-1}}{R_{i+1}-R_{i-1}} \right),
\end{equation}
applied to the radial velocities $v_i$ measured at $R_i$ of the observed rotation curve. Calculations
are performed only where the rotation curve is smooth enough, because of the enormous velocity fluctuations
and low spatial resolution, which cause large dispersions in our results. The estimated shearing
parameters from observational rotation curves are presented in Fig.~\ref{fig:shear}. Different
values of the parameter $q$, correspond to the following interpretations:
	when $q < 0$, the rotation velocity increases faster than a solid body;
	when $q = 0$, we have solid body rotation;
	when $0 < q < 1$, the rotation velocity increases slower than a solid body;
	$q = 1$ relates to a flat rotation curve;
	for $q > 1$, the azimuthal rotation decreases with $R$.
We found that the shearing rates are high in all three galaxies and due to the large variations in
the rotation curves, $q$ changes rapidly. However, in the case of NGC~6822, $q$ gradually increases from
0 to 1.5 with galactocentric distance. For the galaxies IC~10 and NGC~4449, the calculated local
shearing rates vary from $-1.5$ to 3. This scatter in the results is caused by the large fluctuations
in the measured rotation velocities.

\subsection{The magnetic field evolution}

\begin{table*}[t]
\centering
\caption{Summary of the simulations results}
\begin{tabular}{l|r@{.}lr@{.}lr@{.}lr@{.}l|cc|l}
\hline \hline
Model         
	& \multicolumn{2}{c}{$\log\bar{E}_B^{\it end}$} 
		& \multicolumn{2}{c}{$\log E_B^{\it out}$}
			& \multicolumn{2}{c}{$E_B^{\it out}/\bar{E}_B^{\it end}$}
				& \multicolumn{2}{c|}{$\left<B\right>$}
          & $t_e$
            & $T_\Omega$
              & Description \\
        
	& \multicolumn{2}{c}{} 
		& \multicolumn{2}{c}{}
			& \multicolumn{2}{c}{}
				& \multicolumn{2}{c|}{[$\mu$G]}
          & [Myr]
            & [Myr]
              & \\
\hline
R.01Q1$^a$       & 1&37 & 0&07 &  0&05 &  0&068 & 1\,219 & 628 & slow rotation \\
R.02Q1           & 3&55 & 2&02 &  0&03 &  0&835 &    440 & 314 & slow rotation \\
R.03Q1$^b$       & 4&03 & 2&89 &  0&07 &  1&206 &    375 & 209 & medium rotation \\
R.04Q1           & 4&17 & 3&29 &  0&13 &  1&285 &    346 & 157 & fast rotation \\
R.05Q1$^c$       & 3&95 & 3&43 &  0&30 &  1&120 &    509 & 125 & fast rotation \\
\hline
R.01Q0           &-1&73 &-0&42 & 20&48 &  0&001 &     -- & 628 & low shear\\
R.01Q.5          & 0&06 &-0&35 &  0&39 &  0&011 &     -- & 628 & medium shear \\
R.01Q1$^a$       & 1&37 & 0&07 &  0&05 &  0&068 & 1\,219 & 628 & medium shear \\
R.01Q1.5         & 1&26 & 0&15 &  0&08 &  0&056 & 1\,232 & 628 & high shear \\
R.05Q0           &-1&00 & 0&04 & 10&78 &  0&003 &     -- & 125 & low shear \\
R.05Q.5          & 4&03 & 3&38 &  0&22 &  1&168 &    363 & 125 & medium shear \\
R.05Q1$^c$       & 3&95 & 3&43 &  0&30 &  1&120 &    509 & 125 & medium shear \\
R.05Q1.5         & 3&81 & 3&07 &  0&18 &  0&934 &    375 & 125 & high shear \\
\hline
SF3R.03Q.5       & 3&14 & 2&23 &  0&12 &  0&190 &    506 & 209 & low SFR \\
SF3R.03Q1        & 3&28 & 2&28 &  0&10 &  0&243 &    444 & 209 & low SFR \\
SF10R.03Q.5$^d$  & 3&87 & 2&28 &  0&03 &  0&991 &    403 & 209 & medium SFR \\
SF10R.03Q1$^b$   & 4&03 & 2&89 &  0&07 &  1&206 &    375 & 209 & medium SFR \\
SF30R.03Q.5      & 2&79 & 2&77 &  0&96 &  0&190 &    629 & 209 & high SFR \\
SF30R.03Q1       & 3&30 & 3&05 &  0&57 &  0&243 &    547 & 209 & high SFR \\
\hline
M10/100          & 3&78 & 2&73 &  0&09 &  0&833 &    422 & 209 & \\
M20/200$^d$      & 3&87 & 2&28 &  0&03 &  0&243 &    403 & 209 & \\
M50/100          & 4&05 & 2&92 &  0&07 &  1&001 &    404 & 209 & \\
M100/200         & 4&11 & 2&67 &  0&04 &  1&352 &    393 & 209 & \\
M100/100         & 3&86 & 2&72 &  0&07 &  0&751 &    422 & 209 & constantly exploding SNe\\
FIRST            & 2&37 & 1&33 &  0&09 &  0&113 &    572 & 209 & SN activity during first 50Myr \\
\hline
\end{tabular}
\begin{flushleft}
The subsequent columns show: the model name, the mean value of total magnetic energy $\bar{E}_B^{\it
end}$ over past $50~\textrm{Myr}$, the total outflow of magnetic energy during whole simulation (see
Eq.  \ref{eq:outflow}), the ratio of these two quantities, the final mean value of the magnetic
field in a disc midplane $\left<B\right>$ ($|z|<$~$20$~pc), the e-folding time of magnetic flux
increase $t_e$, galaxy revolution timescale $T_\Omega$, and a short description of a model.
Superscripts are explained in Table~\ref{tab:models}. Values of magnetic field energy are normalized
to the initial value.
\end{flushleft}
\label{tab:outflow}
\end{table*}

\begin{figure*}[t]
\centering
\includegraphics[height=0.3\textheight,keepaspectratio=true]{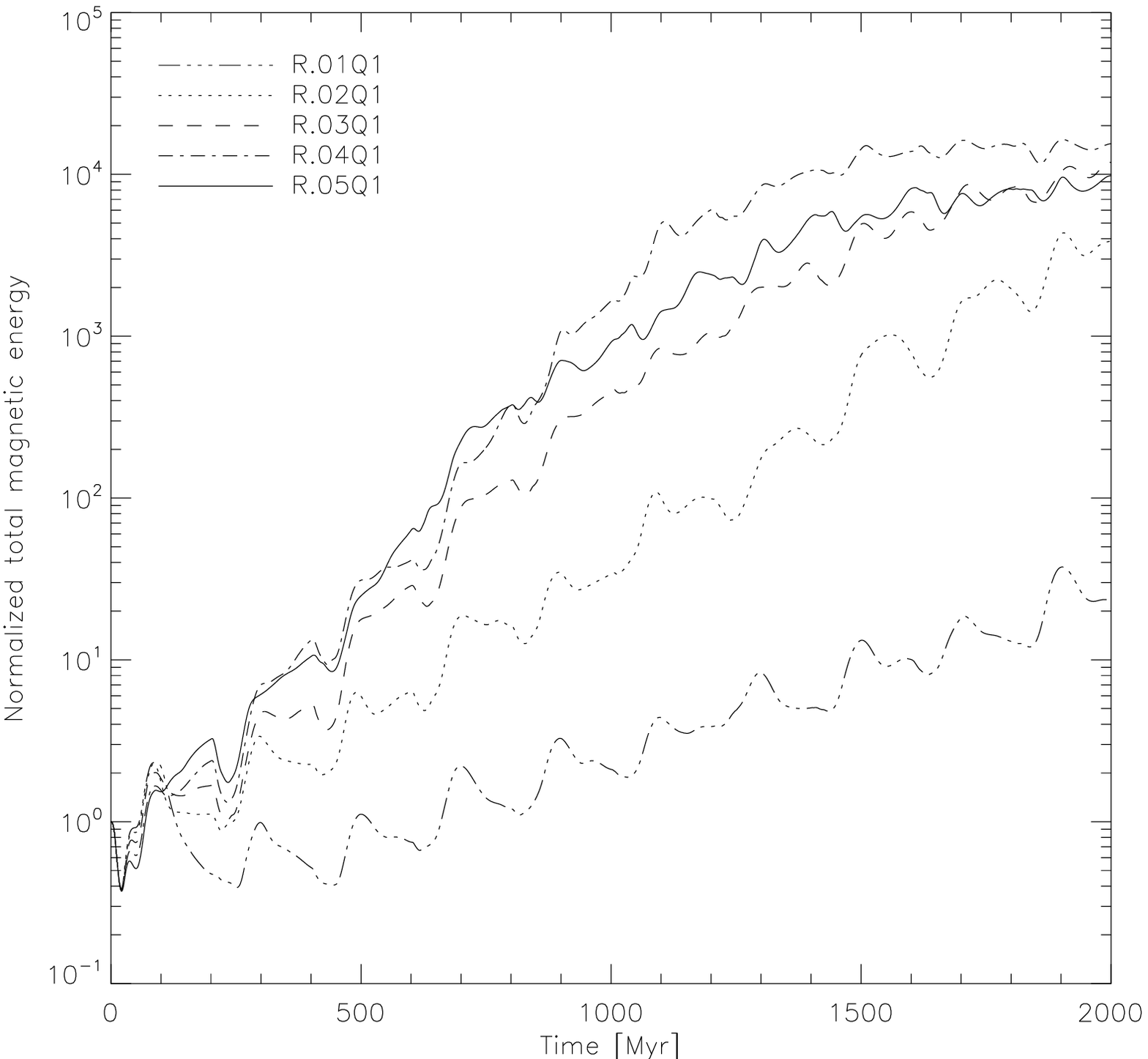}
\hspace{0.5cm}
\includegraphics[height=0.3\textheight,keepaspectratio=true]{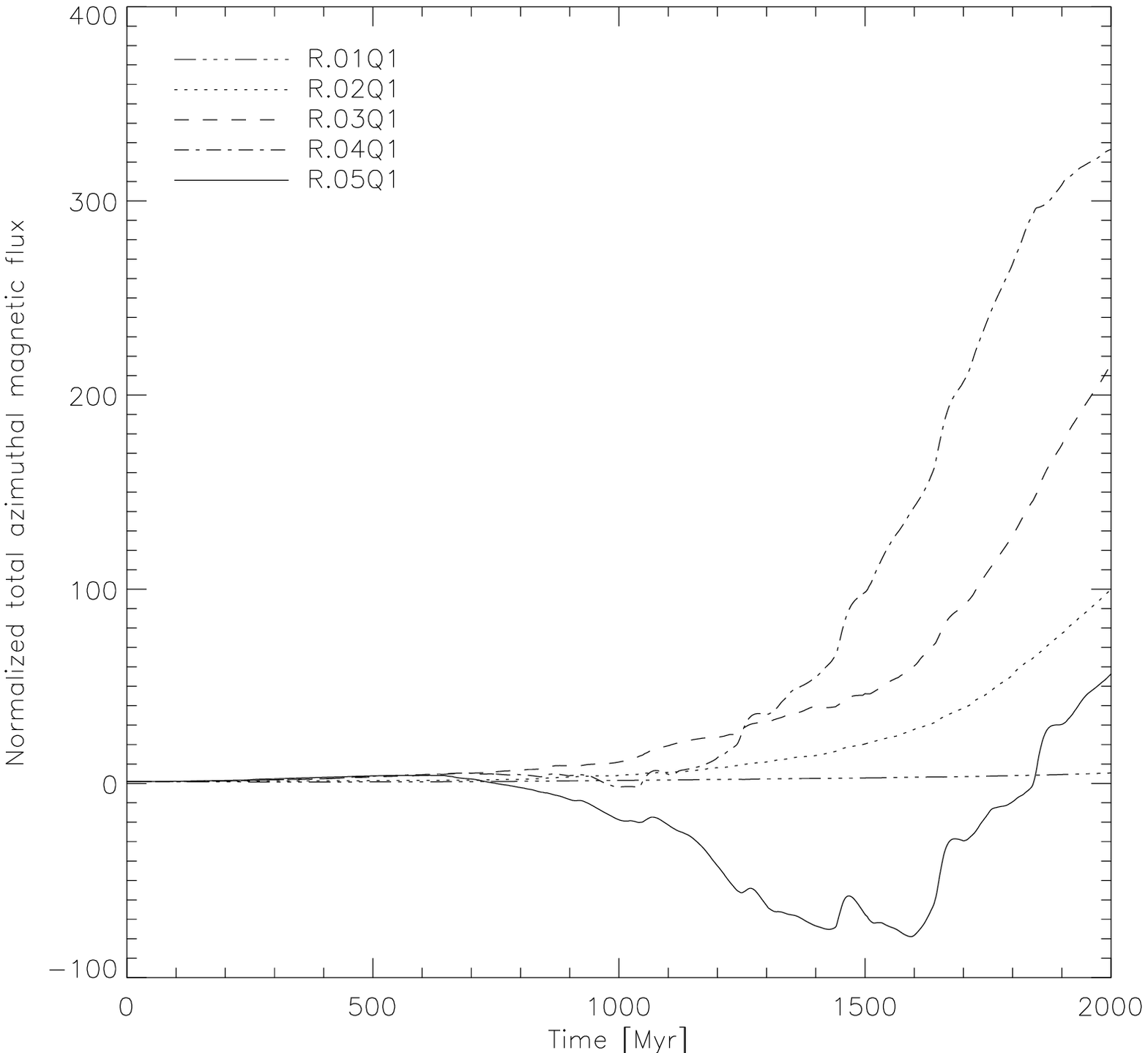}
\caption{Evolution of the total magnetic energy $E_B$ (left panel) and the total azimuthal flux $B_\phi$
(right) for models with different rotation. Both quantities are normalized to the initial value.}
\label{fig:rotation}
\end{figure*}

We study dependence of the magnetic field amplification on the parameters describing the rotation
curve, namely, the shearing rate $q$ and the angular velocity $\Omega$. The evolution in the total
magnetic field energy $E_B$ and total azimuthal flux $B_\phi$ for different values of $\Omega$ is
shown in Fig.~\ref{fig:rotation}, left and right panel, respectively.  Models with higher angular
velocities, starting from 0.03~Myr$^{-1}$, initially exhibit exponential $E_B$ growth and after
about 1\,200~Myr, the process saturates (see Sect.  \ref{sec:discusion} for the discussion). The
saturation values of magnetic energy for these three models are similar and $E_B$ exceeds the value
$10^4$ in the normalized units. The magnetic energy in the models R.01Q1 and R.02Q1 grows
exponentially during the whole simulation and does not reach the saturation level. The final $E_B$
for R.02Q1 is around $4\times10^3$ and for the slowest rotation (R.01Q1) in our sample is only $20$.
The total azimuthal magnetic flux evolution (Fig.~\ref{fig:rotation}, right) shows that a higher
angular velocity produces a higher amplification. The azimuthal flux for models with $\Omega \ge
0.02$~Myr$^{-1}$ exceeds the value $10^2$. Model R.01Q1 does not enhance the azimuthal flux at all.

\begin{figure*}[t]
\centering
\includegraphics[height=0.3\textheight,keepaspectratio=true]{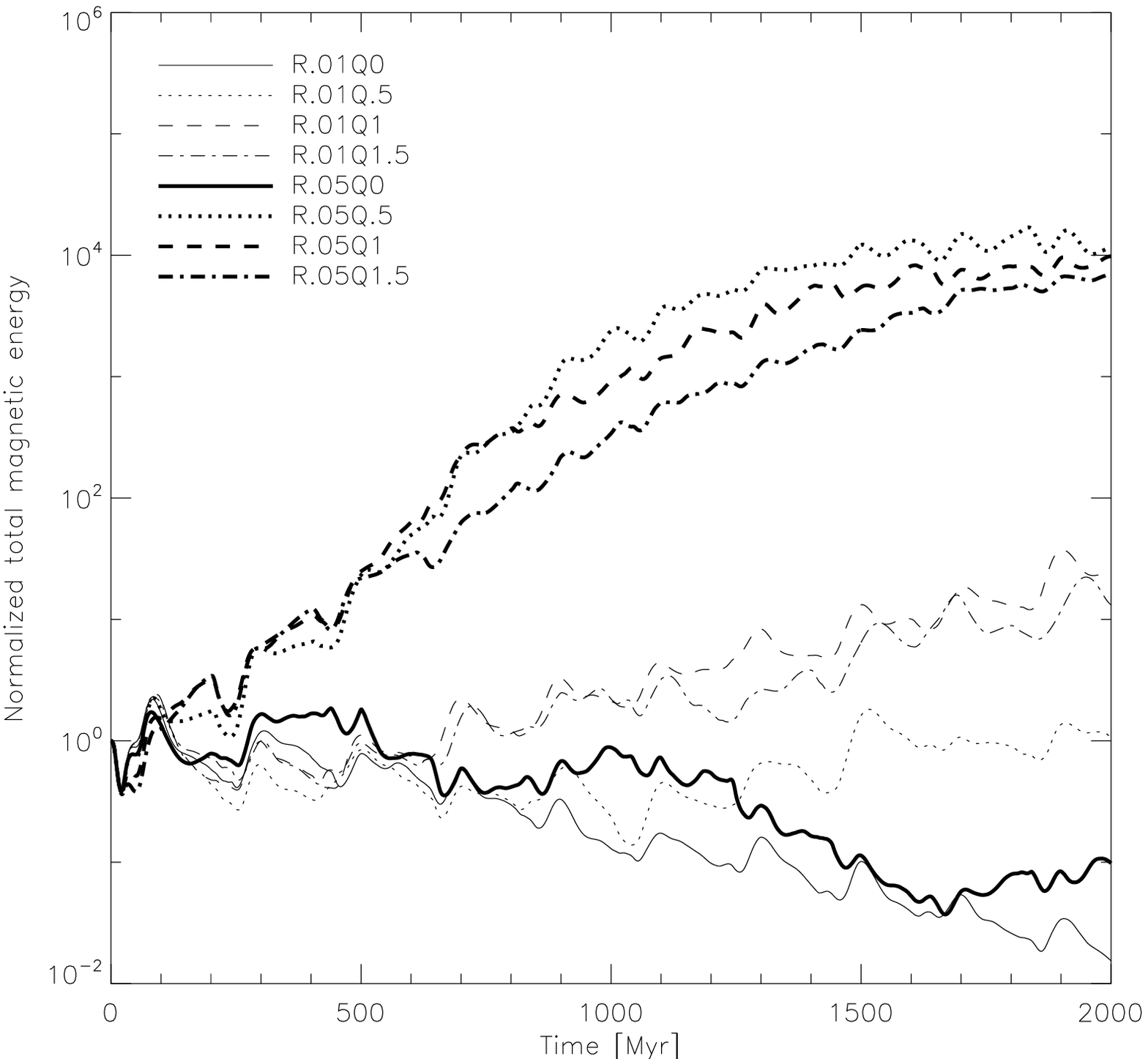}
\hspace{0.5cm}
\includegraphics[height=0.3\textheight,keepaspectratio=true]{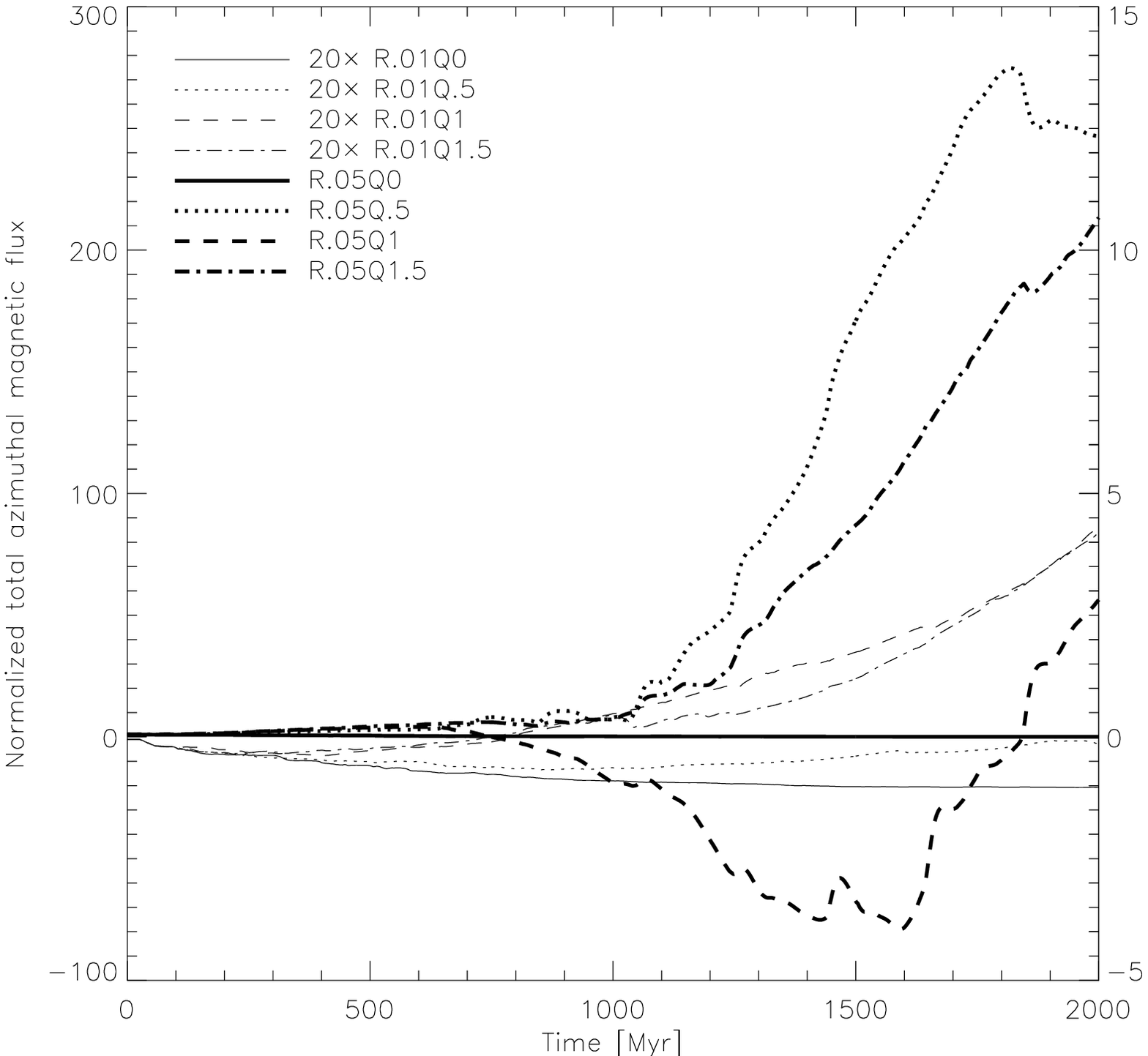}
\caption{Evolution of the total magnetic energy $E_B$ (left panel) and the total azimuthal flux $B_\phi$
(right) for models with different shearing rate and rotation. Both quantities are normalized to the
initial value. For models with $\Omega = 0.01$~Myr$^{-1}$, the $B_\phi$ values has been multiply by
factor 20 -- proper y-axis for these plots is on the right side of the frame.}
\label{fig:shearing}
\end{figure*}

In Fig. \ref{fig:shearing}, we present results for models with different shearing rate values.  The
evolution of $E_B$ and $B_\phi$ in models R.05Q.5 and R.05Q1.5 follows the evolution of model
R.05Q1, which is described in the previous paragraph. Similar behavior is noted for R.01Q1.5 and
R.01Q1, but the model R.01Q.5 alone sustains its initial magnetic field. In the case of models with
no shear (R.01Q0, R.05Q0), the initial magnetic field decays.

\begin{figure*}[t]
\centering
\includegraphics[height=0.3\textheight,keepaspectratio=true]{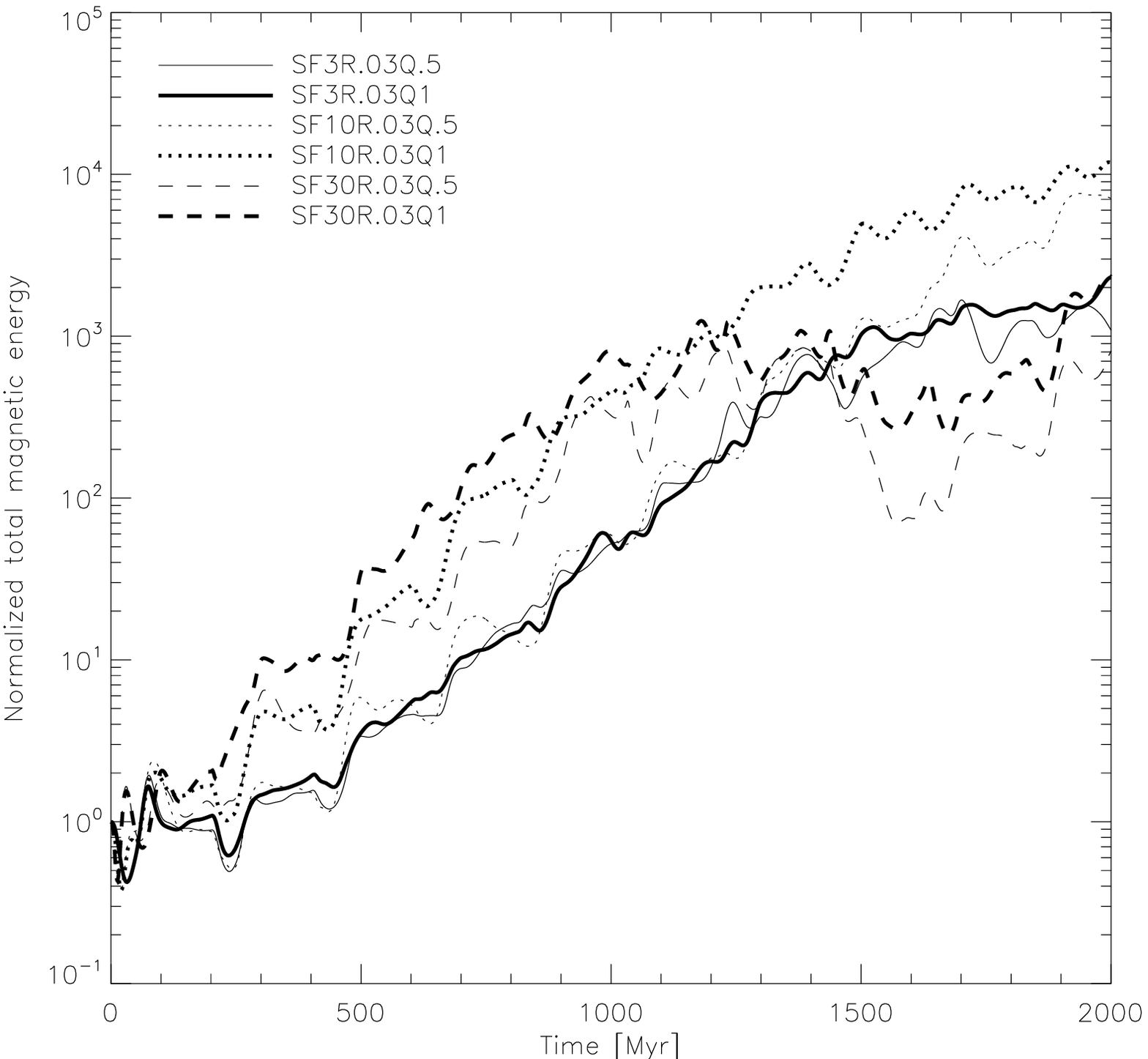}
\hspace{0.5cm}
\includegraphics[height=0.3\textheight,keepaspectratio=true]{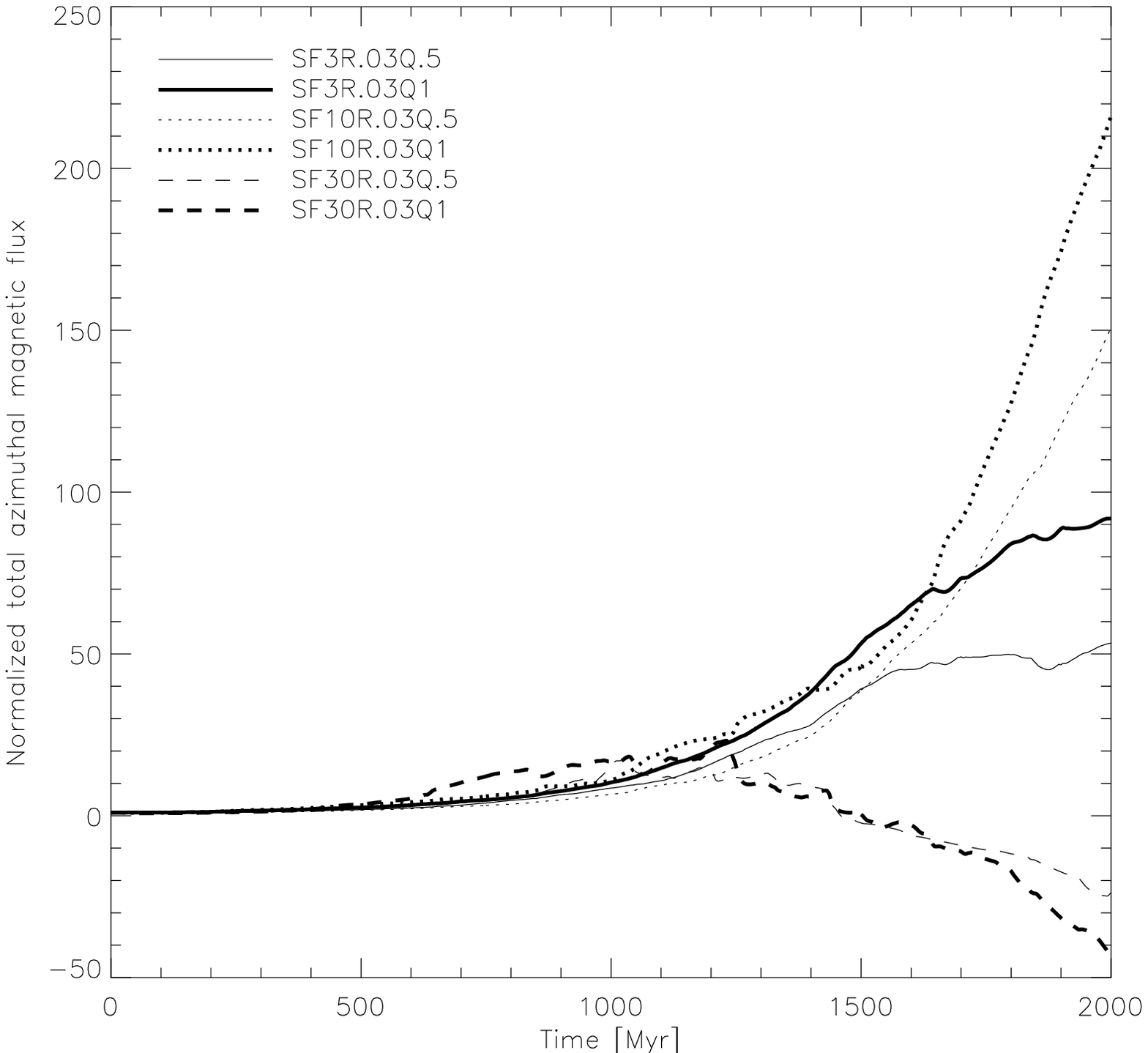}
\caption{Evolution of the total magnetic energy $E_B$ (left panel) and the total azimuthal flux
$B_\phi$ (right) for models with different supernova explosion frequency and shearing rate. Both
quantities are normalized to the initial value.} \label{fig:frequency}
\end{figure*}

\begin{figure*}[t]
\centering
\includegraphics[height=0.3\textheight,keepaspectratio=true]{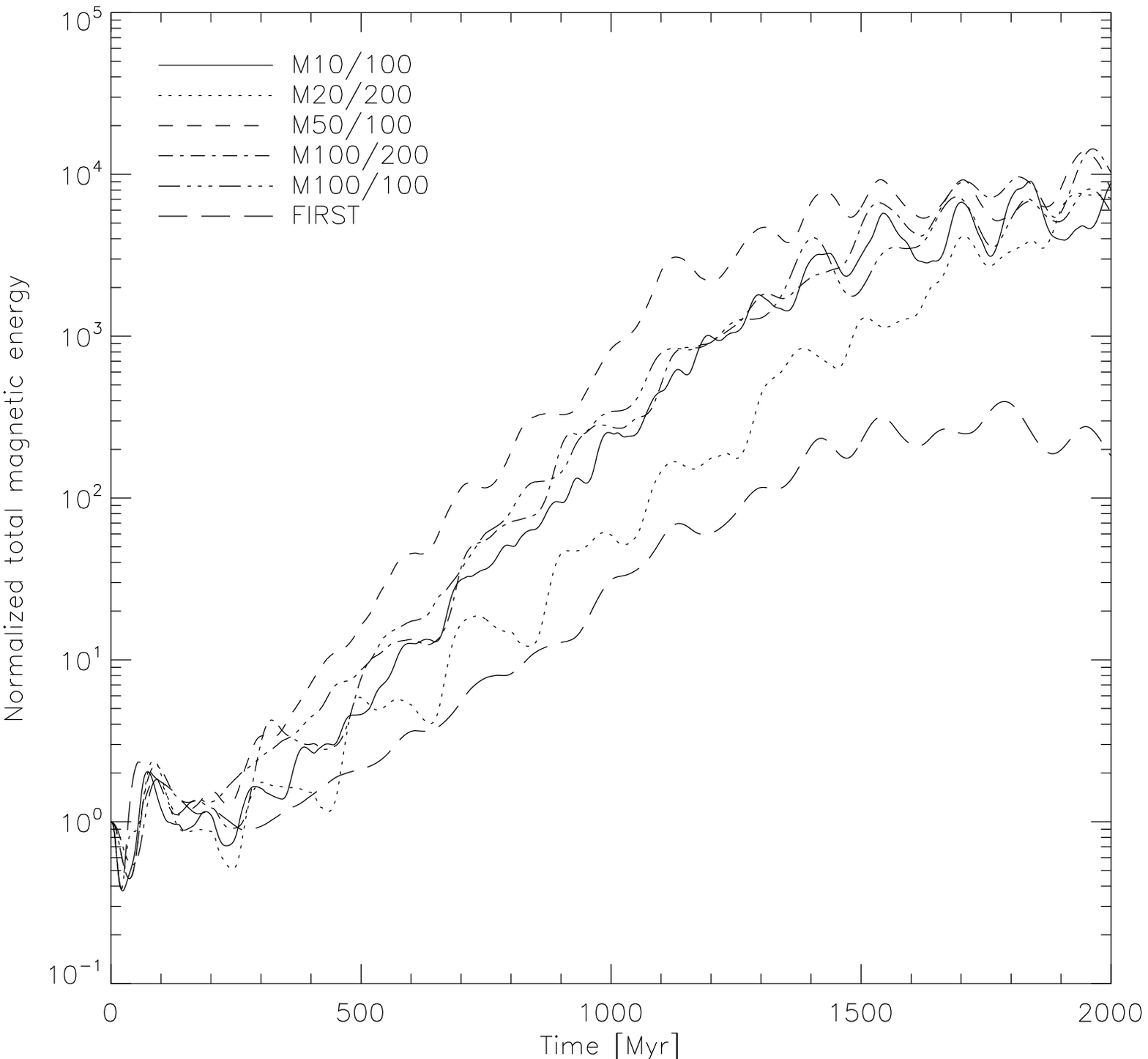}
\hspace{0.5cm}
\includegraphics[height=0.3\textheight,keepaspectratio=true]{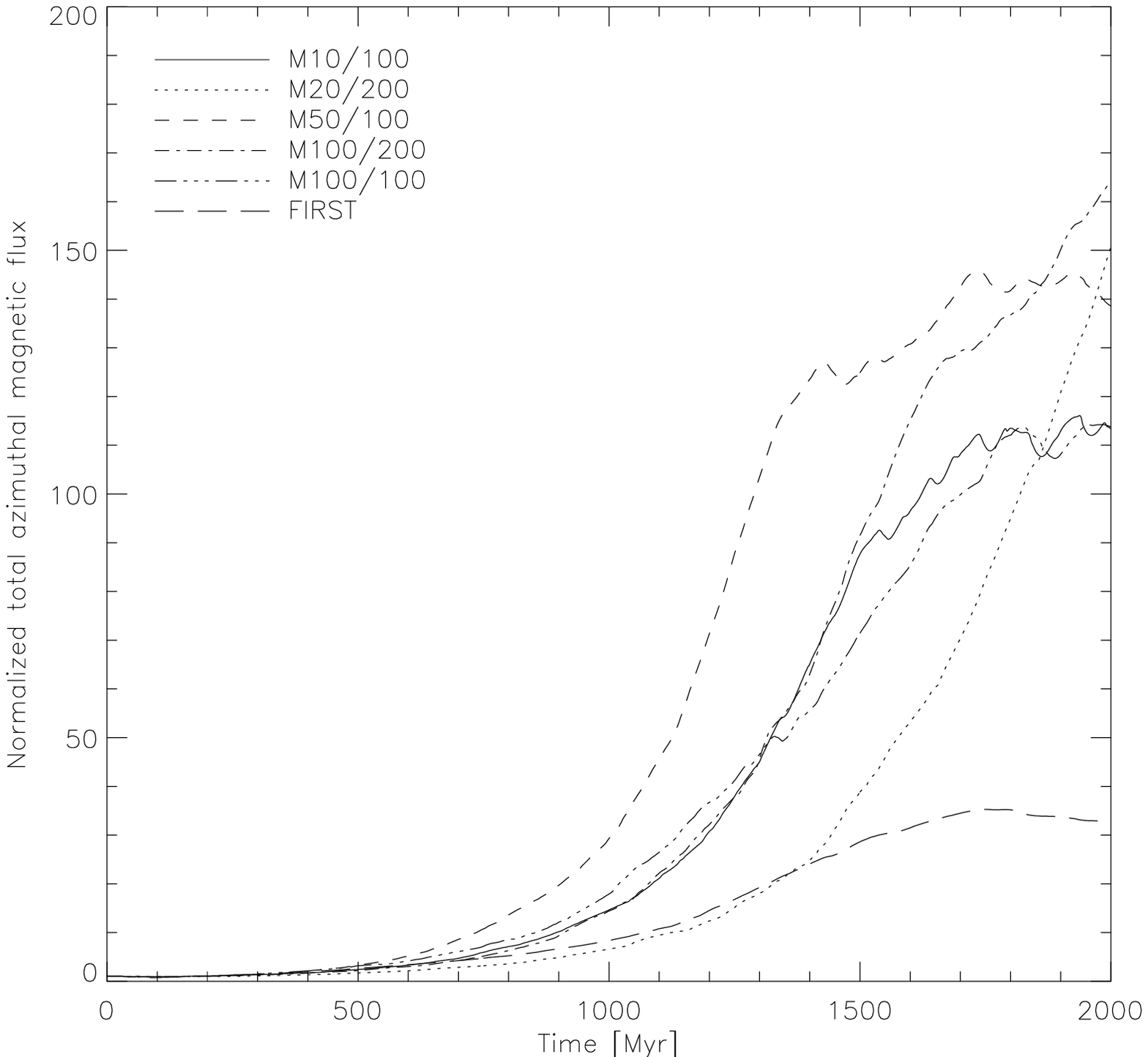}
\caption{Evolution of the total magnetic energy $E_B$ (left panel) and the total azimuthal flux
$B_\phi$ (right) for models with different times of supernovae modulation. Both quantities are
normalized to the initial value.} \label{fig:modulation}
\end{figure*}

We check how the frequency and modulation of SNe influence the amplification of magnetic fields.
The evolution in total magnetic field energy and total azimuthal flux for different supernova
explosion frequencies are shown in Fig. \ref{fig:frequency}, left and right respectively. The total
magnetic energy evolution for all models is similar, but in the case of the azimuthal flux we
observe differences between the models. The most efficient amplification of $B_\phi$ appears for
SF10R.03Q.5 and SF10R.03Q1, and for other models the process is less efficient.  In addition, for
models SF30R.03Q.5 and SF30R.03Q1, we observe a turnover in magnetic field direction. The results
suggest that the dynamo requires higher frequencies of supernova explosions to create more regular
fields, although, if the explosions occur too frequently, this process is suppressed because of the
overlapping turbulence. The analysis of the M models (Fig. \ref{fig:modulation}) shows that the
dynamo process depends on the duration of the phase when supernova activity switches off.  The
fastest growth of magnetic field amplification occurs for models M100/200 and M50/100 in which
periods of SN activity occupy half of the total modulation period. The amplification is apparently
weaker in cases of short SN activity periods (M10/100, M20/200) and continuous activity (M100/100).
In all M models, the final $E_B$ reaches a value of the order of $10^4$.  For $B_\phi$ evolution, we
found that the magnetic flux in the model M50/100 increases exponentially and saturates after
1\,300~Myr. Similar behavior is exhibited by the models M10/100 and M100/100 but the saturation
times occur after 1\,700~Myr and the growth is slower than in the previous case. The model M100/200
after exponential growth at $t=1\,650$~Myr probably begins to saturate, but to quantify this exactly
the simulation should continue. The model M20/200 grows exponentially and does not appear to
saturate.

In the case of the model FIRST (Fig. \ref{fig:modulation}), we found that after about 8~galaxy
revolutions the growth in $E_B$ and $B_\phi$ stops. The total magnetic field energy increase
exponentially and after reaching a~maximum at $t=1\,400$~Myr, it exceeds the value $2\times10^2$,
whereas the azimuthal flux saturates after $1\,600$~Myr and afterwards starts to decay gently.

\subsection{Magnetic field outflow}

To measure the total production rate of magnetic field energy during the simulation time, we calculate
the outflowing $E_B^{\it out}$ through the $xy$ top and bottom domain boundaries. To estimate the
magnetic energy loss, we compute the vertical component of the Poynting vector
\begin{equation}
S_z = (B_x v_x + B_y v_y) B_z - (B_x^2 + B_y^2) v_z.
\end{equation}
This value is computed in every cell belonging to the top and bottom boundary planes and then
integrated over the entire area and time:
\begin{equation}
E_B^{\it out} = \frac{1}{\Delta_z} \sum_{t}\left(
\sum_{ij} (S_z)^t_{ijk_{\it min}} - (S_z)^t_{ijk_{\it max}}
\right)\Delta t,
\label{eq:outflow}
\end{equation}
where $k_{\it min}$ and $k_{\it max}$ refer to the bottom and top boundary respectively, $\Delta_z$
is the vertical dimension of single cell, $t$ is the simulation time, and $\Delta t$ is the
timestep.  For models with a~low dynamo efficiency most of the initial magnetic field energy is
transported out of the simulation box. In some cases (i.e., all models except R.01Q0 and R.05Q0), we
find that the energy loss $E_B^{\it out}$ is comparable to the energy remaining inside the domain
$\bar{E}_B^{\it end}$. In these models, the ratio $E_B^{\it out}/\bar{E}_B^{\it end}$ varies from
0.03 to 0.96 and is highly dependent on the supernova explosion frequency. For models with $q = 0$,
in which the dynamo does not operate, the outflowing energy originate only from the initial
condition for the magnetic field. The results show that the outflowing magnetic energy is
substantial (see Table~\ref{tab:outflow}) suggesting, that irregular galaxies because of their
weaker gravity can be efficient sources of intergalactic magnetic fields.

\section{Discussion} \label{sec:discusion}

The most effective magnetic field amplification that we have found is that in the model R.04Q1,
which we associate with the galaxy NGC~4449. This galaxy has the highest star formation rate in our
sample of three irregulars. The rotation is rapid, reaching 40\,km/s, and, for a wide range of
radii, the shear is strong.  The numerical model predicts an effective magnetic field amplification
and NGC~4449 indeed hosts the strongest magnetic field among the irregulars, both in terms of its
total and ordered component of 14~$\mu$G and 8~$\mu$G, respectively (Chy\.zy et al.~2000).

The next galaxy IC~10 forms stars at a lower rate than NGC~4449. The shear is strong and it is a
rapid rotator. We can compare this galaxy to our model R.05Q1, where we see the fastest initial
growth of the total magnetic energy, but the final value is smaller than that in the case of slower
rotation.  The total azimuthal flux evolves in a complex way with a reversal in the mean magnetic
field direction.  This may indicate that because of its relatively rapid rotation and small size,
instabilities can evolve faster. Separate instability domains can mix (overlap) with each other
resulting in a chaotic though still amplified magnetic field. Consequently IC~10 exhibits a strong
total magnetic field of 5--15\,$\mu$G (as estimated by Chy\.zy et al. 2003). We notice that by
increasing the rotation speed, the amount of magnetic energy expelled from the galaxy grows (see
Table~\ref{tab:outflow}). IC~10 has a~relatively low mass and its shallow gravitational potential
makes the escape of its magnetized ISM easier.

NGC~6822 forms stars at the slowest rate in our sample. It is also the slowest rotator.  The
rotation is almost rigid in its central part (out to $\sim$0.5\,kpc) gradual becoming differential
at larger galactocentric distances but the calculated shearing rate remains small.  We can explain
its weak magnetic field of lower than 5\,$\mu$G (Chy\.zy et al.~2003) by comparing with our model
FIRST: a~single burst of star-forming activity in the past followed by a long (lasting until
present) period of almost no star-forming activity.  In this model, the magnetic field, amplified
initially, fades since the star formation stops. This star-forming activity was analyzed for spiral
galaxies by Hanasz et al.  (2006), who measured a~linear growth in the magnetic field.  We can
explain this by using a shorter simulation time (by about a factor of two) than in our case, but it
may indicate that in irregulars the magnetic field is more easily expelled from the galaxy. 

Our models, for which we measure amplification in $E_B$ and $B_\phi$ during the simulation, produce
a mean magnetic field of order 1--0.5~$\mu$G  (Table~\ref{tab:outflow}) within a disc volume. Models
with slower growth of magnetic field reach values of $\left<B\right>$ around tens of nG, and models
with no dynamo action diffuse the initial magnetic field outside the simulation box.

In Table~\ref{tab:outflow}, we present the average e-folding time of the magnetic flux increase
$t_e$ and the galactic revolution period $T_\Omega$. The $t_e$ of most models is in between 300 and
600~Myr.  For spiral galaxies, Hanasz et al. (2006, 2009) found that the e-folding timescale is
about 150--190~Myr. The difference between spirals and irregulars is probably caused by rotation,
which is much more rapid in spirals.

In most of our models, large fractions of the magnetic field are expelled out of the computational
domain -- almost 20\%--30\% of the magnetic energy maintained in the galaxy. In general more rapid
rotation and a high SNe rate make it easier for the magnetized medium to escape. However, for higher
shear rate, the share of the expelled magnetic field is lower. The optimal set of parameters, from
this point of view, is represented by the model R.05Q1, which we relate to IC\,10. In the other two
galaxies, the expelled field is also high -- about 10\%. Models with excessive star formation
increase this fraction to 60\% (SF30R.03Q.5) or even 96\% (SF30R.03Q1). Therefore, the irregular
galaxies, in particular compact and intensively forming stars such as IC\,10, are an important
source of magnetic field in intergalactic and intracluster media, as predicted by Kronberg et al.
(1999). 

For most of our models we found that the value of the magnetic field strength in the vicinity of a
galaxy (at $z=4$~kpc) is about 30-200~nG. Only models with low magnetic-field production rates
produce negligible magnetic fields at this height. This area is the highest point in our simulation
domain above the galactic midplane and can be considered as a transition region between the ISM and
the IGM. Hence, the magnetic field strengths in the models can be an upper limit to the values in
the IGM region. Our estimates are in an agreement with previous studies, including Ryu et al.
(1998), who demonstrated that in largescale filaments, magnetic fields of about 1~$\mu$G may exists,
Kronberg et al. (1999), who calculated that on Mpc scales the average magnetic field strength is
about 5~nG, and Gopal-Krishna \& Wiita (2001), who showed that radio galaxies can seed the IGM with
a magnetic field of the order 10~nG during the quasar era. However, to obtain realistic profile or
even the maximum possible range of expelled magnetic field in the case of dwarf galaxies we should
take into account the interaction between the IGM and ISM (pressure), which is not included in our
model. We plan to extend our research in this respect in future work.
 
\section{Conclusions}
We have described the evolution in the magnetic fields of irregular galaxies in terms of a
cosmic-ray driven dynamo. Our cosmic-ray driven dynamo model consists of (1) randomly exploding
supernovae that supply the CR density energy, (2) shearing motions due to differential rotation, and
(3) ISM resistivity. We have studied the amplification of magnetic fields under different conditions
characterized by the rotation curve (the angular velocity and the shear) and the supernovae activity
(its frequency and modulation) typical of irregular galaxies.  We have found that:
\begin{itemize}
\item in the presence of slow rotation and weak shear in irregular galaxies, the amplification
of the total magnetic field energy is still possible;
\item shear is necessary for magnetic field amplification, but the amplification itself depends
weakly on the shearing rate;
\item higher angular velocity enables a higher efficiency in the CR-driven dynamo process;
\item the efficiency of the dynamo process increases with SNe activity, but excessive SNe
activity reduces the amplification;
\item a shorter period of halted SNe activity leads to faster growth and an earlier saturation time in the
evolution of azimuthal magnetic flux;
\item for high SNe activity and rapid rotation, the azimuthal flux reverses its direction because
of turbulence overlapping;
\item because of the shallow gravitation potential of an irregular galaxy, the outflow of magnetic
field from the disk is high, suggesting that they may magnetize the intergalactic medium as
predicted by Kronberg et al.  (1999) and Bertone et al. (2006).
\end{itemize}

The performed simulations indicate that the CR-driven dynamo can explain the observed magnetic
fields in irregular galaxies. In future work we plan to determine the influence of other ISM
parameters and perform more global simulations of these galaxies.

\acknowledgements{
This work was supported by Polish Ministry of Science and Higher Education through grants:
92/N-ASTROSIM/2008/0 and 3033/B/H03/2008/35. Presented computations have been performed on the
GALERA supercomputer in TASK Academic Computer Centre in Gda\'nsk.}


\begin{thebibliography}{}
\bibitem{} Bajaja, E., Huchtmeier, W.K., Klein, U. 1994, A\&A, 285, 285
\bibitem{} Beck, R. 2005, in The Magnetized Plasma in Galaxy Evolution, Eds. K.~Chy\.zy,
K.~Otmianowska-Mazur, M.~Soida, and R.-J.~Dettmar, Krak\'{o}w, p. 193
\bibitem{} Berezinskii, V.S., Bulanov, S.V., Dogiel, V.A., Ptuskin, V.S. 1990, Astrophysics of
cosmic rays, ed. V.L. Ginzburg (Amsterdam: North-Holland)
\bibitem{} Bertone, S., Vogt, C., En\ss{}lin, T. 2006, MNRAS, 370, 319
\bibitem{} Chy\.{z}y, K.T., Beck, R., Kohle, S., Klein, U., Urbanik, M. 2000, A\&A, 355, 128
\bibitem{} Chy\.{z}y, K.T., Knapik, J., Bomans, D.J., Klein, U., Beck, R., Soida, M., Urbanik,~M.
2003, A\&A, 405, 513
\bibitem{} Ferri\`{e}re, K. 1998, ApJ, 497, 759
\bibitem{} Everett, J.E., Zweibel, E.G., Benjamin, R.A., McCammon, D., Rocks, L., Gallagher, J.S.
2008, ApJ, 674, 258
\bibitem{} Gaensler, B.M., Haverkorn, M., Staveley-Smith, L., Dickey, J.M., McClure-Griffiths, N.M.,
Dickel, J.R., Wolleben, M. 2005, Science, 307, 1610
\bibitem{} Gallagher, J.S., Hunter, D.A. 1984, ARA\&A, 22, 37
\bibitem{} Giacalone, J., Jokipii, R.J. 1999, ApJ, 520, 204
\bibitem{} Gopal-Krishna, Wiita, P.J. 2001, ApJ, 560, L115
\bibitem{} Gressel, O., Elstner, D., Ziegler, U., R\"udiger, G. 2008, A\&A, 486, L35
\bibitem{} Hanasz, M., Otmianowska-Mazur, K., Lesch, H. 2002, A\&A, 386, 347
\bibitem{} Hanasz, M., Lesch, H. 2003, A\&A, 404, 389 
\bibitem{} Hanasz, M., Kowal, G., Otmianowska-Mazur, K., Lesch, H. 2004, ApJ, 605, L33
\bibitem{} Hanasz, M., Kowal, G., Otmianowska-Mazur, K., Lesch, H. 2006, AN, 327, 469
\bibitem{} Hanasz, M., Otmianowska-Mazur , K., Kowal, G., Lesch, H. 2009, A\&A 498, 335 
\bibitem{} Hawley, J.F., Gammie, C.F., Balbus, S.A. 1995, ApJ, 440, 742
\bibitem{} Hunter, D.A., Wilcots, E.M.; van Woerden, H.; Gallagher, J. S.; Kohle, S. 1998, ApJ, 495, 47
\bibitem{} Hunter, D.A., van Woerden, H., Gallagher, J.S. 1999, AJ, 118, 2184
\bibitem{} Jokipii, J.R. 1999, in Interstellar Turbulence, (Cambridge Univ. Press), 70
\bibitem{} Karachentsev, I.D., Karachentseva, V.E., Huchtmeier, W.K., Makarov, D.I. 2004, AJ, 127, 2031
\bibitem{} Kepley, A.A., Muehle, S., Wilcots, E.M., Everett, J., Zweibel, E., Robishaw, T., Heiles,
C. 2007, arXiv:0708.3405
\bibitem{} Klein, U., Haynes, R.F., Wielebinski, R., Meinert, D. 1993, A\&A, 271, 402
\bibitem{} Kronberg, P. P., Lesch, H., Hopp, U. 1999 ApJ, 551, 56
\bibitem{} Lesch, H. 1993, in The Cosmic Dynamo, ed. F. Krause, K.-H. R\"adler \& G. R\"udiger
(Dordrecht: Kluwer), IAU Symp. 157, 395
\bibitem{} Lisenfeld, U., Wilding, T.W., Pooley, G.G., Alexander, P. 2004, MNRAS, 349, 1335
\bibitem{} Luks, Th., Rohlfs, K. 1992, A\&A, 263, 41
\bibitem{} Martin, C.L. 1998, ApJ, 506, 222
\bibitem{} Otmianowska-Mazur, K., Chy\.zy, K.T., Soida, M., von Linden, S. 2000, A\&A, 359, 29
\bibitem{} Ryu, D., Kim, J., Hong S.S., Jones, T.W. 2003, ApJ, 668, 338
\bibitem{} Ryu, D., Kang, H., Biermann, P.L. 1998, A\&A, 335, 19
\bibitem{} Schlickeiser, R., Lerche, I. 1985, A\&A, 151, 151
\bibitem{} Strong, A.W., Moskalenko, I.V., Ptuskin, V.S., 2007, Annu. Rev. Nucl. Part. S.
\bibitem{} Vallenari, A., Bomans, D.J. 1996, A\&A, 313, 713
\bibitem{} Valdez-Guti\'errez, M., Rosado, M., Puerari, I., Georgiev, L., Borissova, J.,
Ambrocio-Cruz, P. 2002, ApJ, 124, 3157
\bibitem{} Weldrake, D.T.F, de Blok, W.J.G., Walter, F. 2003, MNRAS, 340, 12
\bibitem{} Widrow, L.M. 2002, RvMP, 74 775
\bibitem{} Wilcots, E.M., Miller, B.W, 1998, ApJ, 116, 2363

\end{thebibliography}
\end{document}